\documentclass{article}
\usepackage{spconf,amsmath,graphicx}
\usepackage{caption, subcaption} 
\usepackage{color}
\usepackage{cite}
\usepackage{url}
\usepackage{float}
\usepackage{booktabs}
\usepackage{multirow}
\usepackage{tabularx}
\PassOptionsToPackage{hyphens}{url}\usepackage{hyperref}
\usepackage[skip=-1pt]{caption}
\title{Don\textquotesingle t speak too fast:  The impact of data bias \\ on self-supervised speech models}
\name{Yen Meng$^{\star \dagger}$*, Yi-Hui Chou$^{\dagger}$*, Andy T. Liu$^{\star \dagger}$, Hung-yi Lee$^{\star \dagger}$ \thanks{*Equal Contribution}}
\address{
$^{\star}$Graduate Institute of Communication Engineering, National Taiwan University\\
$^{\dagger}$College of Electrical Engineering and Computer Science, National Taiwan University\\
\{r10942085, b06901012, f07942089, hungyilee\}@ntu.edu.tw}

\begin{document}
\ninept
\maketitle
\begin{abstract}
Self-supervised Speech Models (S3Ms) have been proven successful in many speech downstream tasks, like ASR. However, how pre-training data affects S3Ms' downstream behavior remains an unexplored issue.
In this paper, we study how pre-training data affects S3Ms by pre-training models on biased datasets targeting different factors of speech, including gender, content, and prosody, and evaluate these pre-trained S3Ms on selected downstream tasks in SUPERB Benchmark. Our experiments show that S3Ms have tolerance toward gender bias.  Moreover, we find that the content of speech has little impact on the performance of S3Ms across downstream tasks, but S3Ms do show a preference toward a slower speech rate.
\end{abstract}
\section{Introduction}
\label{sec:intro}
Self-supervised Learning (SSL) from raw speech has become increasingly popular in recent studies, as SSL achieves state-of-the-art results on various downstream tasks \cite{nguyen20_interspeech, 9414460, Huang2020SelfsupervisedPR}, ranging from speaker identification, automatic speech recognition, intent classification and a lot more.  Recent work also dive into the interpretability of Self-supervised Speech Models (S3Ms), as models can be sensitive to powerful adversarial attacks.  Researchers are curious about what these models have really learned, and most work focus on the explainability of model mechanisms or learned representations~\cite{Arik_Pfister_2021, yang2020understanding, chung2021similarity}. Nevertheless, how pre-training data affects S3Ms is less studied.

Self-supervised pre-training in most works use only standard datasets, for example, LibriSpeech~\cite{librispeech}, with carefully collected, gender-balanced, and clean data. While it is easy to collect a large amount of unlabeled data, collecting "balanced" data is relatively hard in practical applications. Besides the most discussed gender bias, speech data can also be biased in other aspects, such as content and prosody. However, to our best knowledge, how bias in speech affects SSL is yet unknown. 

In this paper, we present an empirical study on the effects of data bias on S3Ms at the pre-training stage by creating various datasets.  The performance of S3Ms on different pre-training data is evaluated by selected tasks from SUPERB Benchmark \cite{yang21c_interspeech}. Our study in three aspects of data bias provides the following insights :
\begin{enumerate}
    \item \textbf{Gender}: We pre-train S3Ms on datasets with different gender distributions. Our study shows that gender-balanced data is not necessarily needed for effective pre-training.
    \item \textbf{Content}: We pre-train S3Ms on two groups of data with extremely biased speech content. We find that content has little effect on the downstream behavior of S3Ms.
    \item \textbf{Prosody}: We pre-train S3Ms on datasets with faster or slower speech rates. Experiments show that S3Ms pre-trained on slower speech rate lead to better performance overall.
\end{enumerate}

\vspace{-0.5em}

\section{Related Work}
\label{sec:related-work}
\vspace{-0.5em}
One of the most often used learning schemes for self-supervised speech models is through reconstruction speech frames.
Here we introduce some of the recently emerged reconstruction methods.
The Autoregressive Predictive Coding (APC) method~\cite{chung2020generative}, is primarily inspired by language models (LM) for text.
The DeCoAR~\cite{decoar} method combines the bidirectionality of ELMo~\cite{elmo} and the autoregressive reconstruction objective of APC~\cite{chung2020generative}.
The work of \cite{apc2, vq_apc, improved_apc} also adopt the autoregressive reconstruction scheme, but in some variation.
The Transformer Encoder Representations from Alteration (TERA)~\cite{tera} method, an improved version of Mockingjay~\cite{mockingjay}, is mainly inspired by masked language models (MLM)~\cite{bert} for text.
The work of \cite{mpc, mpc2, mpe} also adopt variations of the MLM reconstruction schemes.
In this work, we select two methods to represent each scheme for our study, the APC method from the autoregressive family and the TERA method from the MLM family.
We select two models for our study due to space limitations.

Bias and fairness issues in speech are receiving more attention these days. Demographic bias is the most studied.  Recent works analyze the impact of demographic bias as well as mitigating bias on specific tasks including ASR\cite{meyer-etal-2020-artie, 10.1145/3442188.3445893}, speaker recognition\cite{fenu21_interspeech, fenu2021improving}. and speech translation\cite{bentivogli2020gender, gaido2020breeding, gaido2021split}. A large body of research related to data bias evaluates models on a single downstream task, however, data bias on S3Ms and its effect on downstream tasks from diverse categories are not yet explored. A related work analyzing pre-training data of S3Ms is \cite{hsu2021robust}, which investigates the effect of domain shift in SSL.  Our work differs from theirs in the sense that we focus on data bias toward different speech factors at pre-training stage. 

\vspace{-0.5em}

\section{Experimental Setup}
\label{sec:exp-setup}
\vspace{-0.5em}
As we attempt to investigate the impact of pre-training data bias on S3Ms, the settings for fine-tuning, including fine-tuning data and hyperparameters, are all the same, the only difference is pre-training data.

\subsection{Pre-trained Upstream Models}
For our experiments, we consider two of the most representative S3Ms, Transformer Encoder
Representations from Alteration (TERA)~\cite{tera} and Autoregressive predictive coding (APC) \cite{chung2020generative}. 

\label{ssec:model}
\begin{itemize}
    \item \textbf{TERA: }
    As suggested by the TERA paper, we use two of the alterations proposed by the authors: time and frequency, and pre-train models for 200,000 steps with a batch size of 32.

    \item \textbf{APC: }
    As suggested by the APC paper, we train APC for 100 epochs with a batch size of 32 and use ADAM optimizer with an initial learning rate of $10^{-4}$.
\end{itemize}

\subsection{Datasets}
\label{ssec:datasets}
We hope to explore how bias toward gender distribution, content, and prosody can affect S3Ms, therefore, we design various artificial datasets for pre-training and further evaluate models on downstream tasks.  For faster pre-training and fairness settings, we fix our pre-training data to 100 hr in all the experiments.
We use LibriSpeech (LS) 100 hr and 360 hr to design the 100-hr datasets with different biases as below.

\subsubsection{Gender}
\label{sssec:gender}
Here, We pay attention to the behavior of S3Ms when pre-training data is biased toward gender distribution. Thus, we design datasets with male-to-female ratio as 0:10, 1:9, 2:8, 8:2, 9:1, and 10:0 by randomly sampling files from LS 100 hr and 360 hr. These 6 settings are denoted as \emph{All-F}, \emph{9F1M}, \emph{8F2M}, \emph{2F8M}, \emph{1F9M}, and \emph{All-M} respectively.
To better interpret the results, we randomly sample three 100-hr datasets for each of the gender distribution settings above.  
For 5:5 male to female ratio (denoted as \emph{5F5M}), we use the original LS 100 subset\footnote{The original LS 100 subset is gender balanced.} as well as 3 random sampled 100-hr datasets from LS 100 hr plus 360 hr.

\subsubsection{Content}
\label{sssec:content}
In this section, we aim to explore whether pre-training on "complex" or "simple" sentences affects S3Ms' downstream behavior. 
Here we use the perplexity (ppl) of the transcription of an utterance measured from a language model to determine whether a sentence is complex or simple. 
We utilize the LS official ARPA language model to calculate ppl for each transcription in the LS 100 hr and 360 hr subset and create two datasets, 100 hr audio with the highest ppl and 100 hr audio with the lowest ppl, denoted as \emph{ppl high} and \emph{ppl low} respectively. 
Audios in \emph{ppl high} contain rarer words and proper nouns, while audios in \emph{ppl low} are mostly composed of common and simple words.


\subsubsection{Prosody}
\label{sssec:prosody}
In addition to gender and content, prosody is also an essential aspect of speech study. Speech rate is viewed as an important prosodic feature, hence in this section, we design the datasets based on speech rates as below. 
We calculate words per minute (wpm) for each sentence in the LS 100 hr and 360 hr subset using the alignments of utterances and the provided transcriptions.
Similar to the setup in Section \ref{sssec:content}, we create two datasets, 100 hr audio with the highest wpm and 100 hr audio with the lowest wpm, denoted as \emph{wpm high} and \emph{wpm low} respectively. Moreover, to further investigate the impact of extreme speech rates on S3Ms, we create two additional artificial datasets by converting the playback speed of all audio files in LS 100 hr subset two times faster and two times slower without altering the pitch. These two datasets are denoted as \emph{speed 2x} and \emph{speed 0.5x} respectively.



\subsection{Downstream Tasks}
\label{ssec:task}

To evaluate the generalizability and effectiveness of pre-trained models on diverse tasks, we select five tasks, which are solvable with linear downstream models, from SUPERB Benchmark\cite{yang21c_interspeech}.  Tasks are carefully chosen as we wish to analyze the effect of data bias, and linear models serve as the direct indication of the quality of speech representations.

For the selected tasks, we follow the settings in the SUPERB Benchmark. The only difference is that instead of using weighted sum to integrate hidden states from all layers, we directly use the last hidden state learned from S3Ms. Salient performance gap may occur for S3Ms with a larger architecture, but since there are only three layers in both TERA and APC, we observe no huge difference between features with and without weighted sum on downstream performance.  The five tasks can be further categorized into four aspects of speech: content, speaker, semantics, and paralinguistics:
\begin{table}[ht]
\centering
\begin{tabular}{cll} 
\toprule
\multirow{2}{*}{\textbf{Content}} & \multicolumn{2}{p{5.5cm}}{\textbf{Phoneme Recognition, PR}: LibriSpeech\cite{librispeech} is adopted. The evaluation metric is phone error rate(PER).} \\ \noalign{\vskip 1mm} \cline{2-3} \noalign{\vskip 1mm} 
& \multicolumn{2}{p{5.5cm}}{\textbf{Keyword Spotting, KS}: Speech Commands dataset v1.0\cite{warden2018speech} is adopted.  The evaluation metric is accuracy(ACC).}\\
\midrule
\textbf{Speaker} & \multicolumn{2}{p{5.5cm}}{\textbf{Speaker Identiﬁcation, SID}: VoxCeleb1\cite{NAGRANI2020101027} is adopted. The evaluation metric is accuracy(ACC).} \\
\midrule
\textbf{Semantics} & \multicolumn{2}{p{5.5cm}}{\textbf{Intent Classification, IC}: Fluent Speech Commands dataset\cite{lugosch2019speech} is adopted. The evaluation metric is accuracy(ACC).}\\
\midrule
\textbf{Paralinguistics} & \multicolumn{2}{p{5.5cm}}{\textbf{Emotion Recognition, ER}: IEMOCAP\cite{Busso2008IEMOCAPIE} is adopted. The evaluation metric is accuracy(ACC).} \\
\bottomrule
\end{tabular}
\end{table}

\vspace{-0.3em}

\section{RESULTS AND ANALYSIS}
\label{sec:result}

\begin{figure}[!htp]
\centering
\begin{subfigure}{\linewidth}
  \centering
  \includegraphics[width=\linewidth,height=4cm]{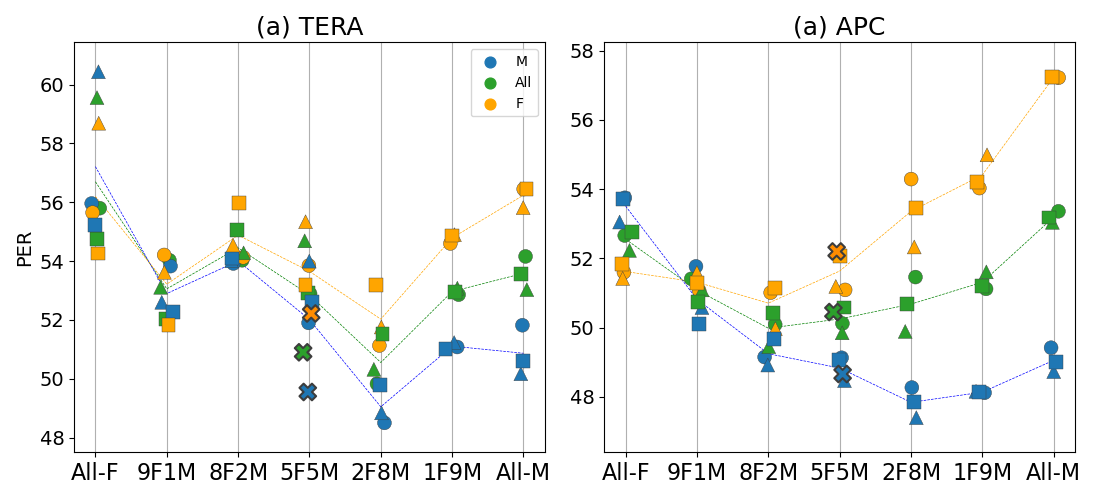}
  \label{fig:pr}
\end{subfigure}
\vspace{-0.5em}
\begin{subfigure}{\linewidth}
  \centering
  \includegraphics[width=\linewidth,height=4cm]{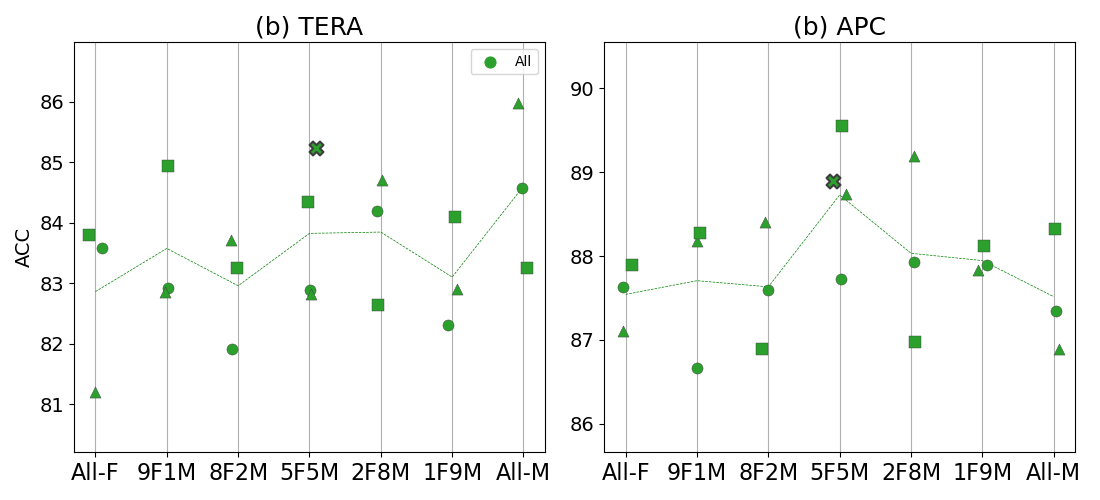}
  \label{fig:ks}
\end{subfigure}
\vspace{-0.5em}
\begin{subfigure}{\linewidth}
  \centering
  \includegraphics[width=\linewidth,height=4cm]{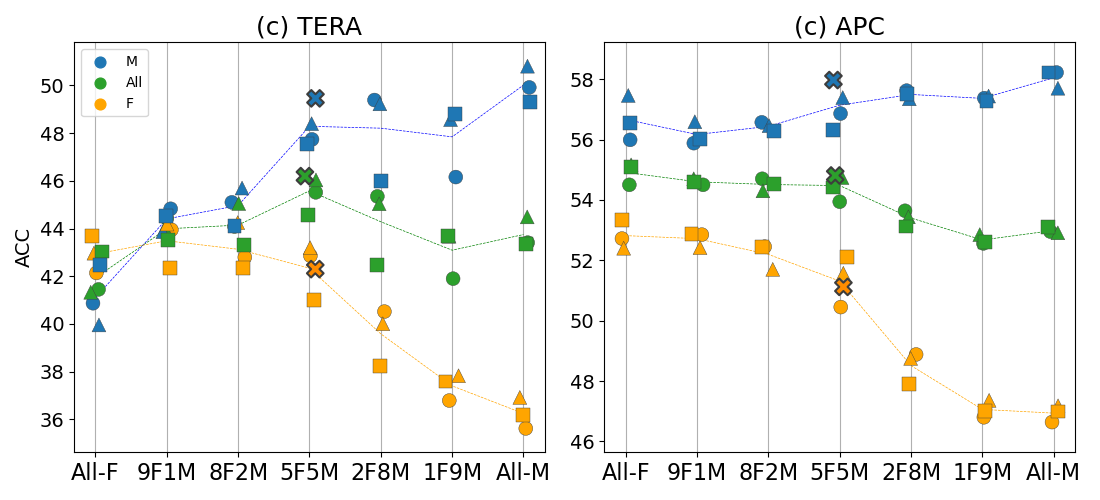}
  \label{fig:sid}
\end{subfigure}
\vspace{-0.5em}
\begin{subfigure}{\linewidth}
  \centering
  \includegraphics[width=\linewidth,height=4cm]{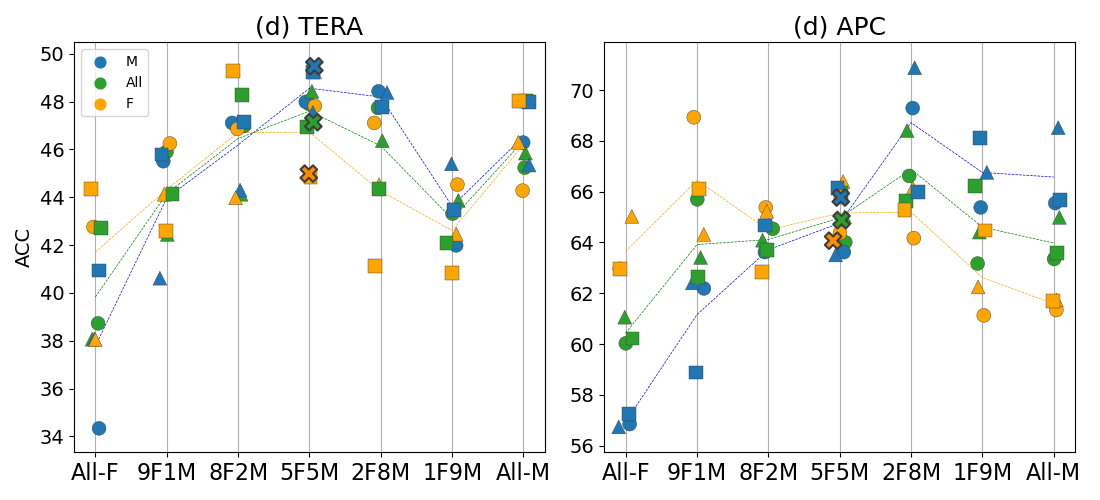}
  \label{fig:ic}
\end{subfigure}
\vspace{-0.5em}
\begin{subfigure}{\linewidth}
  \centering
  \includegraphics[width=\linewidth,height=4cm]{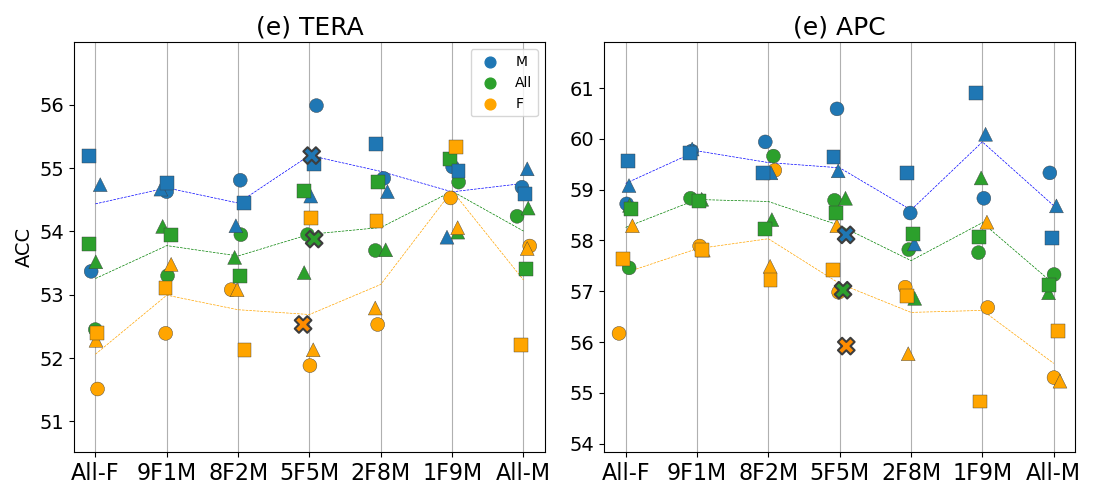}
  \label{fig:er}
\end{subfigure}
\caption{\textbf{(a) PR (b) KS (c) SID (d) IC (e) ER} \newline Results (in \%) of both S3Ms (TERA and APC) pre-trained on data with different gender distribution. Data points come with three different shapes, indicating three random sampled pre-training dataset.  The dashed line is obtained by connecting the average of three data points. In 5F5M, the 'x' marker with bolder outline represents the LS 100 hr dataset. Notation---`M': \texttt{male}, `F': \texttt{female}}.
\label{fig:1}
\end{figure}

\subsection{Gender}
\label{ssec:gender-result}
\subsubsection{Downstream Behavior}
\label{sssec:downstream}
Figure \ref{fig:1} shows that, in general, S3Ms pre-trained on balanced data achieve the best result; however, pre-training models on gender-imbalanced data does not always cause a severe degrade in downstream performance. 
At the most extreme setting, namely \emph{All-F} and \emph{All-M}, most of S3Ms perform the worst.  
But this situation can be mitigated by adding a small amount of data from the other gender, which can effectively elevate the performance, models can then be on par with, or even better than, those pre-trained on balanced data.

The testing sets of selected tasks are approximately gender-balanced. For further investigation on gender bias during testing, we split the testing set by gender and evaluate S3Ms on male and female subsets separately, except for the dataset of KS, where no demographic information is provided.  As such, the overall testing score will be the average of scores on male and female subsets.

We observe that some tasks seem to be affected more by gender bias, such as PR with APC and SID with both S3Ms.  When pre-training models on datasets with higher male voice ratio, the accuracy on female subset drops rapidly, and vice versa.  We conjecture this is because the diversity of female voice is much higher than that of male voice. However, we can bridge the gap between the testing accuracy of the male and female subsets by just adding 10-20 percent of female voice.  For other tasks, gender bias is not obviously presented, for instance, ER is comparatively agnostic to gender bias.

Moreover, figure \ref{fig:1} shows that the behaviors of the models differ across downstream tasks. 
In comparison of the two models, the connected mean line in APC is smoother, while in TERA, the randomness in three random sampled datasets is higher. Still, the overall performance trend in the selected tasks is fairly alike.

\subsubsection{Representation Similarity}
\label{sssec:pwcca}
Following our results in section \ref{sssec:downstream}, we find that the effect on downstream tasks is not as significant as we originally expected, even when gender is highly imbalanced in pre-training data. Hence we are curious whether the representations extracted from S3Ms pre-trained on different gender-biased datasets are similar.
Therefore, we measure the representation similarity of S3Ms pre-trained on gender-biased datasets. 
For similarity measurement, we adopt Projection Weighted Canonical Correlation Analysis (PWCCA) proposed in \cite{pwcca}, and we use LibriSpeech \emph{test-clean} subset. 

\begin{figure}[!ht]
    \centering
    \includegraphics[width=0.95\linewidth]{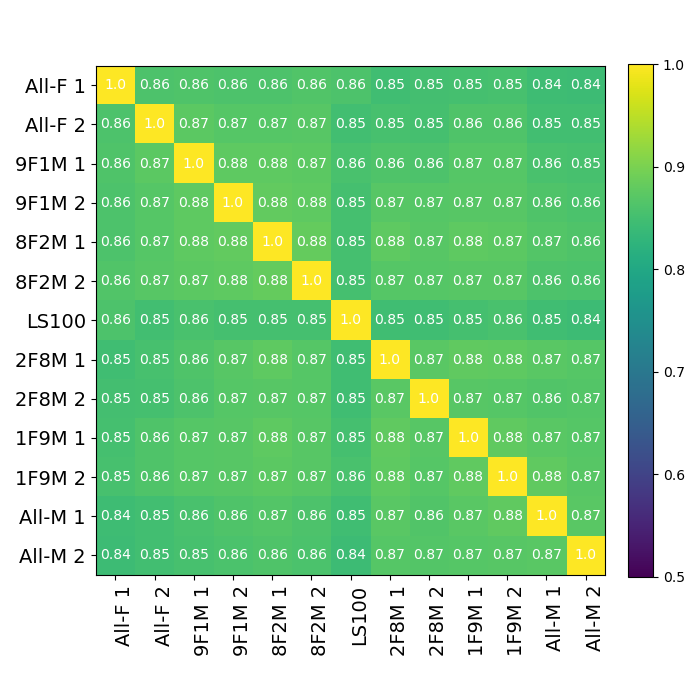}
    \caption{Similarity heatmap among different gender setting in \textbf{TERA}. Similarity values are annotated. The number following gender setting indicates different random sampled pre-training data}
    \label{fig:2}
\end{figure}

\begin{figure}[!ht]
    \centering
    \includegraphics[width=0.95\linewidth]{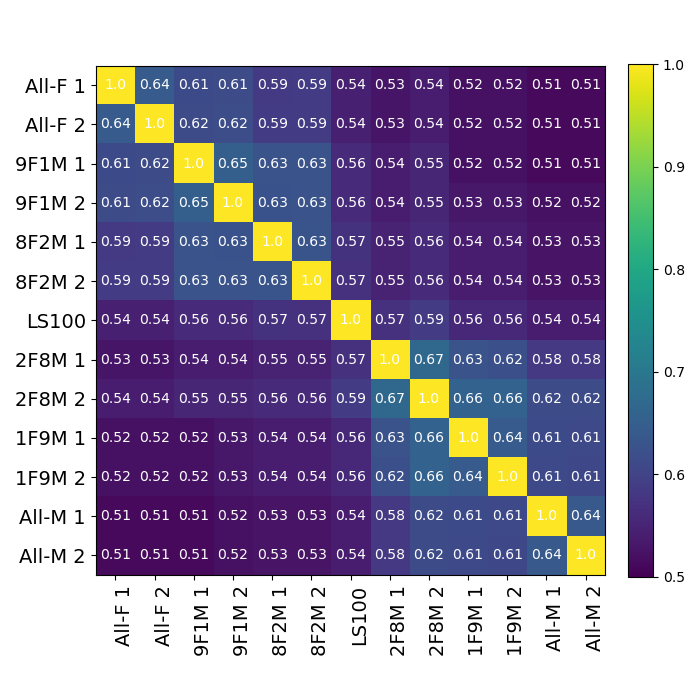}
    \caption{Similarity heatmap among different gender setting in \textbf{APC}. Similarity values are annotated.}
    \label{fig:3}
\end{figure}

From figure \ref{fig:2} and figure \ref{fig:3}, TERA and APC behave differently as we take a closer look at the PWCCA score of their upstream representations. For TERA, the representations of different gender-biased datasets are all very similar, yet we cannot see a higher similarity between two random sampled datasets with the same gender distribution. 
For APC, the overall similarity between different datasets is much lower than that in TERA. However, the upper left corner block and the bottom right corner block show a lighter color, meaning that the representation similarity increases when the gender distribution is more similar. For example, the representation of All-F is much more similar to 9F1M than 1F9M. 
Also, the similarity is the highest between different random sampled datasets under the same gender distribution setting.

While similarity scores for TERA with different gender bias settings are highly alike, we can observe a correlation between gender distribution and its similarity scores for APC. However in both S3Ms, gender-bias in pre-training data has only a slight effect in downstream.
As a result, there is no obvious relationship between representation similarity and small gender bias in downstream tasks.

\vspace{-0.3em}

\vspace{-0.3em}

\begin{table*}[!ht]
\begin{center}
\begin{tabular}{*{12}{l}}
\toprule
&  & \multicolumn{2}{c}{PR} & \multicolumn{2}{c}{KS} & \multicolumn{2}{c}{SID} & \multicolumn{2}{c}{IC} & \multicolumn{2}{c}{ER}\\
& & \multicolumn{2}{c}{PER $\downarrow$} & \multicolumn{2}{c}{ACC $\uparrow$}
& \multicolumn{2}{c}{ACC $\uparrow$} & \multicolumn{2}{c}{ACC $\uparrow$} & \multicolumn{2}{c}{ACC $\uparrow$}\\
\cmidrule(lr){3-4}
\cmidrule(lr){5-6}
\cmidrule(lr){7-8}
\cmidrule(lr){9-10}
\cmidrule(lr){11-12}
& \textbf{pre-train} & \textbf{TERA} & \textbf{APC} & \textbf{TERA} & \textbf{APC} & \textbf{TERA} & \textbf{APC} & \textbf{TERA} & \textbf{APC} & \textbf{TERA} & \textbf{APC} \\
\midrule
Baseline & LS 100  & \textbf{49.64} & 50.45 & 85.23 & 88.90 & \textbf{46.20} & \textbf{56.94} & 47.14 & 64.88 & 54.01 & 56.90\\
\midrule
\multirow{2}{2em}{Content} & ppl high & 51.78 & 50.73 & 83.97 & 88.28 & 42.54 & 54.18 &  44.27 & \textbf{65.67} & 53.85 & \textbf{58.46}\\
& ppl low & 50.94 & \textbf{50.17} & 82.99 & 88.48 & 43.02 & 54.09 & 42.58 & 64.75 & 52.66 & 57.23\\
\midrule
\multirow{4}{4em}{Prosody} & wpm high & 51.60 & 51.97 & \textcolor{red}{81.37} & 87.60 & 44.30 & 54.63 & 44.92 & 62.91 & 53.73 & 57.62\\
& wpm low & 52.38 & 51.10 & \textbf{86.37} & \textbf{89.13} & 43.50 & 53.36 & \textbf{49.93} & 65.12 & 54.36 & 58.21\\
& speed 2x & \textcolor{red}{65.40} & \textcolor{red}{65.47} & 81.73 & \textcolor{red}{83.74} & \textcolor{red}{32.35} & \textcolor{red}{47.55} & \textcolor{red}{35.67} & \textcolor{red}{49.59} & \textcolor{red}{51.89} & \textcolor{red}{54.43}\\
& speed 0.5x & 56.86 & 54.47 & 84.10 & 88.74 & 43.16 & 51.92 & 46.56 & 65.15 & \textbf{54.43} & 57.39\\
\bottomrule
\end{tabular}
\end{center}
\caption{The testing result (in \%) of S3Ms pre-trained on designed datasets. The arrow in the header indicates whether lower/higher score is better. The bold text denotes the best performance on the column, and the red text denotes the worst performance on the column. \emph{Note that for each task, the performance is only compared with the same model.}}
\label{table:1}
\end{table*}

\subsection{Content}
\label{ssec:content}
Table \ref{table:1} lists the testing results of S3Ms (TERA and APC) pre-trained on content and prosody bias.  Surprisingly, we observe that, for both TERA and APC, there is little performance difference between pre-training on either content-biased dataset (\emph{ppl high} and \emph{ppl low}), even evaluated on tasks related to content. For TERA, there is a slight performance drop across five downstream tasks. But for APC, pre-training on content-biased data barely degrades the testing results, and models even outperform the baseline on some tasks such as ER and IC.

\subsection{Prosody}
\label{ssec:prosody}
As table \ref{table:1} shows, the performance difference between \emph{wpm high} and \emph{wpm low} on PR, SID, and ER is not obvious. Nevertheless, we see that data with slower speech rate performs significantly better on KS and IC, especially in TERA.


For \emph{speed 2x} and \emph{speed 0.5x}, a significant difference in downstream performance can be observed. 
Pre-training on \emph{speed 2x} has a considerable performance drop across all tasks.  While using \emph{speed 0.5x} for pre-training slightly degrades performance on PR, KS, and SID, the performance is on par with baseline in terms of IC and even achieves better results on ER.
Interestingly, pre-training S3Ms on data with lower speech rate performs better than data with higher speech rate --- 6 out of 10 even outperform the baseline (LS 100).  Results suggest that it is not too harmful to pre-train on data with extremely slow speech rate, instead, slower speech rate may even be beneficial to some tasks.

\section{Conclusions}
\label{sec:conclusion}
Our work presents an empirical approach for understanding the effect of biased data on S3Ms.  
The quality of speech representations affected by data bias is carefully examined, as we evaluate S3Ms on a wide range of downstream tasks with linear models. Aside from downstream modeling, we also measure representation similarity for more insights, and results show no direct correlation between downstream behavior and representation similarity.
Results on gender bias show that pre-training data does not need to be gender-balanced to ensure the best performance on downstream tasks.  Furthermore, our study suggests that pre-training on biased content does not affect much.  Finally, we find that pre-training S3Ms on data with lower speech rate achieves better performance. 
For future work, the effect of data bias can be studied on more S3Ms from different families. We are also interested to explore other aspects of bias, for instance, single/multiple speakers, synthesized/natural speech, and noisy/quiet environments.

\vspace{-0.5em}

\section{Acknowledgement}
\label{acknowledgement}
\vspace{-0.5em}
We thank to National Center for High-performance Computing (NCHC) of National Applied Research
Laboratories (NARLabs) in Taiwan for providing computational and storage resources.



\bibliographystyle{IEEEbib}
\bibliography{refs}

\end{document}